\documentclass[reprint, amsmath,amssymb, nofootinbib, aps]{revtex4-1}

\usepackage{graphicx}% Include figure files
\usepackage{dcolumn}% Align table columns on decimal point
\usepackage{bm}% bold math
\usepackage{hyperref}% add hypertext capabilities

\usepackage{longtable}
\usepackage[caption=false]{subfig}
\usepackage{array}
\usepackage{color}
\usepackage{colortbl}
\usepackage{multirow}

\providecommand{\U}[1]{\protect\rule{.1in}{.1in}}
\definecolor{Cor1}{gray}{0.95}
\newcolumntype{L}[1]{>{\raggedright\let\newline\\\arraybackslash\hspace{0pt}}m{#1}}
\newcolumntype{C}[1]{>{\centering\let\newline\\\arraybackslash\hspace{0pt}}m{#1}}
\newcolumntype{R}[1]{>{\raggedleft\let\newline\\\arraybackslash\hspace{0pt}}m{#1}}
\ifx\pdfoutput\relax\let\pdfoutput=\undefined\fi
\newcount\msipdfoutput
\ifx\pdfoutput\undefined\else
\ifcase\pdfoutput\else
\ifx\paperwidth\undefined\else
\ifdim\paperheight=0pt\relax\else\pdfpageheight\paperheight\fi
\ifdim\paperwidth=0pt\relax\else\pdfpagewidth\paperwidth\fi
\fi\fi\fi
\begin{document}

\title{Non-linear effects on radiation propagation around a charged compact object}% Force line breaks with \\

\author{R. R. Cuzinatto$^{1}$} \email{rodrigo.cuzinatto@unifal-mg.edu.br}
\author{C. A. M. de Melo$^{1,2}$} \email{cassius.anderson@gmail.com}
\author{K. C. de Vasconcelos$^{3}$} \email{cavalcanti.kelder@gmail.com}
\author{L. G. Medeiros$^{3}$} \email{leogmedeiros@ect.ufrn.br}
\author{P. J. Pompeia$^{4,5}$} \email{pedropjp@ifi.cta.br}

\affiliation{
$^{1}$ Instituto de Ci\^{e}ncia e Tecnologia, Universidade Federal de
Alfenas. Rod. Jos\'{e} Aur\'{e}lio Vilela (BR 267), Km 533,
n${{}^{\circ}} $11999, CEP 37701-970, Po\c{c}os de Caldas, MG, Brazil.
\\ $^{2}$ Instituto de F\'{\i}sica Te\'{o}rica, Universidade Estadual
Paulista. Rua Bento Teobaldo Ferraz 271 Bloco II, P.O. Box 70532-2,
CEP 01156-970, S\~{a}o Paulo, SP, Brazil.\\
$^{3}$ Departamento de F\'{\i}sica Te\'{o}rica e Experimental and
Escola de Ci\^{e}ncia e Tecnologia, Universidade
Federal do Rio Grande do Norte. Campus
Universit\'{a}rio, s/n - Lagoa Nova, CEP 59078-970, Natal, RN, Brazil. \\ $^{4}%
$Instituto de Fomento e Coordena\c{c}\~{a}o Industrial, Departamento
de Ci\^{e}ncia e Tecnologia Aeroespacial. Pra\c{c}a Mal. Eduardo
Gomes 50, CEP 12228-901, S\~{a}o Jos\'{e} dos Campos, SP, Brazil.
\\ $^{5}$ Instituto Tecnol\'{o}gico de Aerona\'{u}tica, Departamento
de Ci\^{e}ncia e Tecnologia Aeroespacial. Pra\c{c}a Mal. Eduardo
Gomes 50, CEP 12228-900, S\~{a}o Jos\'{e} dos Campos, SP, Brazil.
}


\begin{abstract}
The propagation of non-linear electromagnetic waves is carefully analyzed on a curved spacetime created by static spherically symmetric mass and charge distribution. We compute how non-linear electrodynamics affects the geodesic deviation and the redshift of photons propagating near this massive charged object. In the first order approximation, the effects of electromagnetic self-interaction can be distinguished from the usual Reissner-Nordstr\"om terms. In the particular case of Euler-Heisenberg effective Lagrangian, we find that these self-interaction effects might be important near extremal compact charged objects.
\begin{description}

\item[Key-words] Non-linear electrodynamics; Black holes; Redshift;
Light bending; Geodesic path.

\item[PACS numbers] 04.20.Jb,04.70.Bw,41.20.Jb

\end{description}
\end{abstract}

\maketitle

\section{Introduction}

Generalizations of Maxwell electrodynamics have been proposed since
it was established and they are motivated by several reasons such as
experimental constraints on the eventual photon mass
\citep{tu2005mass,cuzinatto2011probe,bonin2010podolsky}, classical
aspects of vacuum polarization
\citep{heisenberg1936folgerungen,schwinger1951}, electrodynamics in
the context of strings and superstrings
\citep{seiberg1999stringnoncom,fradkin1985non,bergshoeff1987born,metsaev1987born,leigh1989dirac},
etc. Among the several generalizations, there is a group known as
Nonlinear Electrodynamics (NLED) which is characterized by presenting
nonlinear field equations. Examples of NLED are Born-Infeld theory
\citep{born1934royal,born1939poincare,born1934foundations,
Stehle1966QEDCorrespondence} and
Euler-Heisenberg electrodynamics \citep{heisenberg1936folgerungen}.
The former was proposed to limit the maximum value of the electric
field of a point charge \citep{delphenich2003nonqed} and the last arises as an effective action
of one-loop QED \citep{dunne2005heisenberg}.

Since the decade of 1980, several applications of NLED in the context of gravitation were proposed
\citep{plebanski1984type,demianski1986static,oliveira1994non,
novello2002analogbh,barcelo2005analoggravity,
cai2008shearviscosity},
including applications to cosmology
\citep{garcia2000born,delorenci2002nonlinear,dyadichev2002non,
moniz2002quintessence,novello2004nonlinear,novello2009cyclic,medeiros2012realistic}
and spherically symmetric solutions of charged Black Holes (BH)
\citep{ayon1998regular,yajima2001black,bronnikov2001regular,diaz2010electrostatic,
diaz2010asymptotically,diaz2013charged,ruffini2013einstein}.
Moreover, generalizations of Reissner-Nordstr\"om solution with NLED
were studied where stability and thermodynamics properties of the BH
were analyzed
\citep{chemissany2008thermodynamics,gunasekaran2012extended,Hendi2012,
diaz2013thermodynamic,Jie2014,breton2014stability,Hendi2014,Hendi2015,
Hendi2015a}.
In particular, the geodesic motion of test particles around
Born-Infeld BH was studied in \citet{linares2014test} and references
therein. However, in most of the papers found in the literature the
direct influence of the background electric field on radiation
propagation is ignored.

In the early 1970's it was realized that the self-interaction of
NLED modifies the dispersion relation of a photon propagating in a
region with a background electric field
\citep{bialynicka1970nonlinear}. From the geometrical point of view,
this modification can be mapped to an effective metric in the flat
spacetime
\citep{boillat1970nonlinear,novello2000geometrical,
novello2010book,DeLorenci2001417,Novello2012Gordonmetric}.
This effective metric can be promptly generalized to a
curved spacetime metric through the minimal coupling prescription.
As a consequence, there are two simultaneous effects on
radiation propagation: first, the path of light rays may be
altered due to self-interaction of the electric field via the
effective metric; second, the photon dynamics is affected by the
spacetime curvature. In fact, some preliminary results
concerning photon propagation in non-linear interaction with a
background electric field and in the presence of a BH were obtained
in the context of Euler-Heisenberg and Born-Infeld electrodynamics
\citep{de2001dyadosphere,breton2002geodesic}. In the present work, the
authors intend to generalize these results for a generic NLED and
particularly show that, despite the fact that the background
electric field does not generate effective horizons (horizons that
would be sensed only by radiation), this field directly influences
the geometric redshift and geodesic deviation.

The paper is organized as follows. In Sect. 2, the spherically
symmetric solution for a generic NLED is determined. In Sect. 3,
minimal coupling prescription is used to generalize the effective
metric in flat spacetime to a curved one. In this section, it is
also shown that the effective metric does not produce any new
horizon. In Sect. 4, the influence of the background electric field
in geometric redshift and geodesic deviation is analyzed. Final
remarks are presented in Sect. 5.

\section{General solution for NLED}

The matter Lagrangian for a general NLED invariant under parity is:
\begin{equation}
\mathcal{L}=\mathcal{L}(F,G^{2}) \, ,
\end{equation}
where $F\equiv-\frac{1}{4}F^{\mu\nu }F_{\mu\nu}$ and
$G\equiv-\frac{1}{4}\tilde{F}^{\mu\nu}F_{\mu\nu}$. The quantity
$F_{\mu\nu}$ is the electromagnetic field tensor and
$\widetilde{F}^{\mu\nu}$ is its dual:
$\widetilde{F}^{\mu\nu}\equiv\frac{1}{2}\eta^{\mu\nu\rho\sigma}F_{\rho\sigma}$,
where $\eta^{\mu
\nu\rho\sigma}=\frac{\varepsilon^{\mu\nu\rho\sigma}}{\sqrt{-g}}$ is
the totally antisymmetric tensor constructed from the Levi-Civita
symbol $\varepsilon^{\mu\nu\rho\sigma}$ and the determinant $g$ of
the metric tensor $g_{\mu\nu}$.\footnote{In flat spacetime,
$F = (  \vec{E}^{2}-\vec{B}^{2} ) / 2 $ and
$G=\vec{E}\cdot\vec{B}$. The vectors $\vec{E}$ and $\vec{B}$ are the
electric and magnetic fields respectively. F and G are the only Lorentz and gauge invariant functions of $F_{\mu\nu}$ \citep{delphenich2006optanalog}.}

The energy-momentum tensor is given by:
\begin{equation}
T_{\mu\nu}=\mathcal{L}_{F}F_{\mu\lambda}^{\text{ \ }}F_{\text{ \ }\nu
}^{\lambda}+G\mathcal{L}_{G}g_{\mu\nu}-\mathcal{L}g_{\mu\nu},\label{tensor em}%
\end{equation}
with $\mathcal{L}_{F}\equiv\frac{\partial\mathcal{L}}{\partial F}$
and $\mathcal{L}_{G}\equiv\frac{\partial\mathcal{L}}{\partial G}$.

The static spherically symmetric line element is:
\begin{equation}
ds^{2}=e^{\nu\left(  r\right)  }dt^{2}-e^{\lambda\left(  r\right)  }%
dr^{2}-r^{2}d\theta^{2}-r^{2}\sin^{2}\theta d\phi^{2}~,\label{ds}%
\end{equation}
where $\nu\left(  r\right)  $ and $\lambda\left(  r\right)  $ are
functions determined by the gravitational field equations. In order to preserve spacetime symmetry,
 one supposes the matter content to be constituted by a static and spherically symmetric electric charge. Hence, only the components $F_{01}%
=-F_{10}=-E\left(  r\right)  $ of the field strength $F_{\mu\nu}$
are non-null. Consequently $G=0$. In this case, the asymptotically
flat exterior Schwarzchild-de Sitter-NLED solution of Einstein
equations is:
\begin{equation}
\nu\left(  r\right)  =-\lambda\left(  r\right)  =\ln\left(  1-\frac{2m}%
{r}+\frac{1}{r}S\left(  r\right)  -\frac{\Lambda}{3}r^{2}\right)  .\label{nu}%
\end{equation}
We consider $c=G_{N}=1$. The constant $m$ is the geometric mass,
$\Lambda$ is the cosmological constant and
\begin{equation}
S\left(  r\right)  \equiv2\int r^{2}\left(  -\mathcal{L}_{F}E^{2}+\mathcal{L}\right)
dr.\label{S(r)}%
\end{equation}

The field equations for $F_{\mu\nu}$ are:
\begin{equation}
\partial_{\nu}\left( \mathcal{L}_{F}\sqrt{-g}F^{\mu\nu}\right)=0 \text{ \ \ and \ \ } \partial_{[\rho}F_{\mu\nu]}=0.\label{EqMaxwell}%
\end{equation}
This implies:
\begin{equation}
\mathcal{L}_{F}E=\frac{q}{r^{2}},\label{E}%
\end{equation}
where $q$ is the total electric charge.

Note that an explicit solution can be obtained when an specific $\mathcal{L}%
(F,G^{2})$ is chosen. For instance, Maxwell Electrodynamics is recovered when $\mathcal{L}_{F}=1$ and
\begin{equation}
E=\frac{q}{r^{2}}\Rightarrow S\left( r\right) =\frac{q^{2}}{r}%
,\label{S_Maxwell}%
\end{equation}
which is the well known Reissner-Nordstr\"om solution.

\section{Effective metric}

In NLED, the electromagnetic field is self-interacting and this is
directly reflected on photon propagation. In particular, when
radiation propagates in a slowly varying electromagnetic field as a
background, the self-interaction between the radiation and the
background field can be described by an effective metric:
$\bar{\eta}^{\mu\nu}$
\citep{bialynicka1970nonlinear,boillat1970nonlinear,novello2000geometrical}.
In flat spacetime one has
\begin{equation}
\bar{\eta}^{\mu\nu}k_{\mu}k_{\nu}=\left( \eta^{\mu\nu}+\lambda_{\pm}%
F^{\mu\beta}F_{\beta}^{\text{ \ }\nu}\right) k_{\mu}k_{\nu}%
=0,\label{metricaefetiva}%
\end{equation}
where
\begin{equation}
\lambda_{\pm} = \frac{-\mathcal{L}_{F}\left( \mathcal{L}_{GG}+\mathcal{L}%
_{FF}\right) +2F\bar{\mathcal{L}} \pm\sqrt{\delta}}{2\left[ G^{2}\bar{\mathcal{L}} +2\mathcal{L}_{F}\left(
\mathcal{L}_{GG}F-\mathcal{L}_{FG}G\right) -\mathcal{L}_{F}^{2}\right]
},\label{z}%
\end{equation}
with
\begin{eqnarray}
\delta = \left[ 2F\left( \mathcal{L}_{FF}\mathcal{L}_{GG}-\mathcal{L}_{FG}%
^{2}\right) +\mathcal{L}_{F}\left( \mathcal{L}_{GG}-\mathcal{L}_{FF}\right)
\right] ^{2} \nonumber \\
+\left[ 2G\left( \mathcal{L}_{FF}\mathcal{L}_{GG}-\mathcal{L}%
_{FG}^{2}\right) -2\mathcal{L}_{F}\mathcal{L}_{FG}\right] ^{2}
\end{eqnarray}
and
\begin{equation}
\bar{\mathcal{L}}=\mathcal{L}_{FF}\mathcal{L}_{GG}-\mathcal{L}_{FG}^{2}. %
\end{equation}
The signal $\pm$ in Eq.~(\ref{z}) indicates birefringence
\citep{akmansoy2013thermodynamics,bialynicka1970nonlinear}, i.e. the
possible dependence of the photon propagation on its polarization,
which is a typical phenomenon of NLED. When minimal coupling
prescription ($\eta_{\mu\nu}\rightarrow g_{\mu\nu}$) is performed,
the effective metric on the curved spacetime is found to be:
\begin{equation}
\bar{g}^{\mu\nu}=g^{\mu\nu}+\lambda_{\pm}F^{\mu\beta}F_{\beta}^{\text{ \ }\nu
}, \label{g bar}
\end{equation}
where $g^{\mu\nu}$ is the spacetime metric given in Eq.~(\ref{ds})
-- see appendix A of \citet{novello2000geometrical} for more
details. This way, photons (i.e. weak radiation propagating fields
that do not disturb the spacetime) will follow geodesics described
by the following effective line element
\begin{equation}
d\bar{s}^{2} = \frac{e^{\nu}}{\left[ 1+\lambda_{\pm}E^{2} \right]} dt^{2} - \frac{e^{-\nu
}}{\left[ 1+\lambda_{\pm}E^{2}\right]}dr^{2} -r^{2}d\Omega^{2}.\label{metrica cov}%
\end{equation}
This line element contains both spacetime (gravitational) effects --
\emph{via} the factor $e^{\nu}$ -- and the influence of the
background field -- through the function
$\left[1+\lambda_{\pm}E^{2}\right] $. The angular part of the line
element is $d\Omega^2 = d\theta^{2} +\sin^{2}\theta d\phi^{2}$, as
usual. It is also worth mentioning that the NLED under consideration
influences photon trajectory both gravitationally -- through
$S\left( r\right) $ -- and electromagnetically -- by means of
$\lambda_{\pm}$. For instance, in Maxwell electrodynamics:
$\lambda_{\pm}=0$ and $S\left( r\right) $ is given by
Eq.~(\ref{S_Maxwell}). On the other hand, in Born-Infeld theory,
where
\begin{equation}
\mathcal{L}_{\text{BI}}=b^{2}\left[ 1-\sqrt{1-\frac{2F}{b^{2}}-\frac{G^{2}}{b^{4}}%
}\right] , \label{L_BI}%
\end{equation}
one finds $\lambda_{\pm}=\frac{1}{b^{2}-E^{2}}$; $S\left(
r\right) $ is given by elliptic functions
\citep{breton2002geodesic}. Note that, both in Maxwell and Born-Infeld cases, no
birefringence is present \citep{boillat1970nonlinear,demelo2014causal}.

The first point to be analyzed here concerns the existence of
effective event horizons. For this goal, an investigation of the radial
null geodesics is carried on using Eq.~(\ref{metrica cov}). The result is:
\begin{equation}
\left( \frac{dt}{dr}\right) ^{2}=\frac{\left[ 1+\lambda_{\pm}E^{2}\right]
e^{-\nu}}{\left[ 1+\lambda_{\pm}E^{2}\right] e^{\nu}}=e^{-2\nu}, \label{dt/dr}
\end{equation}
which is exactly the same expression obtained considering
propagation with the spacetime metric Eq.~(\ref{ds}). This means
that the horizons ``seen'' by photons are generated exclusively by
gravitational effects of the spacetime. Note that this result is
formally independent of the particular type of NLED under
consideration because the factors scaling with $\lambda_{\pm}$ cancel
out in Eq. (\ref{dt/dr}). However, we emphasize that the
gravitational effects of the NLED are still present in $\nu$.

Although the horizons do not depend on the effective metric, the
later is important in other phenomena such as geometric redshift and
light deflection. These two effects are analyzed in the following
section.

\section{Influence of NLED on radiation propagation}

In this section, an investigation on how an specific NLED influences
the radiation redshift and geodesic deviation will be conducted. No
influence of the cosmological constant will be considered here, i.e.
$\Lambda=0$. A perturbative approach will be adopted, in which
nonlinear effects are small corrections to the Maxwell theory.
$\mathcal{L}\left( F,G^{2}\right) $ is written as
\begin{equation}
\mathcal{L}\left( F,G^{2}\right) \simeq F+\frac{1}{2}a_{+}F^{2}+\frac{1}%
{2}a_{-}G^{2}+\mathcal{O}\left( 3\right) \label{Lagrangena aproximada}%
\end{equation}
where $\mathcal{O}\left( 3\right) $ represents higher order terms
and
\begin{equation}
a_{+}F\sim a_{-}G\ll1.\label{aproxim1}%
\end{equation}
The condition $a_{\pm}>0$\emph{ }ensures that the energy density is
positive definite \citep{diaz2010asymptotically}.

In the first order regime of perturbation and considering a radial
electric field ($F=\frac{E^{2}}{2}$ e $G=0$), Eq.~(\ref{z}) reduces to:
\begin{align}
\lambda_{+}  &  \simeq\frac{\mathcal{L}_{FF}}{\mathcal{L}_{F}}\simeq
a_{+}\left( 1-\frac{a_{+}}{2}E^{2}\right) , \nonumber \\
\lambda_{-}  &  \simeq\frac{\mathcal{L}_{GG}}{\mathcal{L}_{F}-2F\mathcal{L}%
_{GG}}\simeq a_{-}\left( 1-\frac{a_{+}}{2}E^{2}+a_{-}E^{2}\right) .
\end{align}
Hence,
\begin{equation}
1+\lambda_{\pm}E^{2}\simeq1+a_{\pm}E^{2}.
\end{equation}

The next step is to obtain $S\left( r\right) $. By substituting
Eq.~(\ref{Lagrangena aproximada}) into Eq.~(\ref{E}), one obtains:
\begin{equation}
\left( 1+\frac{a_{+}}{2}E^{2}\right) E\simeq\frac{q}{r^{2}},
\end{equation}
which leads to:
\begin{equation}
E\simeq\frac{q}{r^{2}}\left[ 1-\frac{a_{+}}{2}\left( \frac{q}{r^{2}}\right)
^{2}\right] .\label{Sol_E}%
\end{equation}
Notice that the condition (\ref{aproxim1}) implies:
\begin{equation}
a_{+}\left( \frac{q}{r^{2}}\right) ^{2}\ll1.
\end{equation}

Now Eq.~(\ref{Sol_E}) is substituted in Eq.~(\ref{S(r)}) leading to
\begin{equation}
S\left( r\right) \simeq\frac{q^{2}}{r}-\frac{k}{20}\frac{q^{4}}{r^{5}},
\end{equation}
where $k \equiv a_{+}$. The change of notation $k$ instead of $a_{+}$ is
done in order to distinguish gravitational effects of the NLED from
purely electromagnetic effects, i.e. $k$ is related to the spacetime
metric and $a_{\pm}$ to the effective metric.

Finally, the components of the metric Eq.~(\ref{ds}) in the first
order approximation are rewritten as:
\begin{align}
\bar{g}_{00}  &  = \frac{e^{\nu}}{\left[ 1+\lambda_{\pm}E^{2}\right]}\simeq
W\left( r\right)\left[1-a_{\pm}\left( \frac{q}%
{r^{2}}\right) ^{2} \right]-\frac{k}{20}\frac{q^{4}}{r^{6}},\label{g00}\\
\frac{-1}{\bar{g}_{11}}  &  =\frac{ \left[ 1+\lambda_{\pm}E^{2}\right] }{ e^{-\nu} }
\simeq W\left( r\right)\left[1+a_{\pm}\left( \frac{q}%
{r^{2}}\right) ^{2} \right] -\frac{k}{20}\frac{q^{4}}{r^{6}},\label{-g11}%
\end{align}
where
\begin{equation}
W\left( r\right) =1-\frac{2m}{r}+\frac{q^{2}}{r^{2}}\label{W}%
\end{equation}
is the usual Reissner-Nordstr\"om term.

In this approximation, the manner by which the event horizon is
affected by the NLED can be studied. The horizons $r_{\pm}$ are
obtained when the condition $g^{11}=0$ is considered, which in our
particular case gives:
\begin{equation}
1-\frac{2m}{r}+\frac{q^{2}}{r^{2}}-\frac{k}{20}\frac{q^{4}}{r^{6}}%
\simeq0.\label{Eq_horizontes}%
\end{equation}
The approximated solution to this equation is:
\begin{equation}
r_{\pm}\simeq r_{\pm(0)}\pm\frac{r_{\pm(0)}^{2}}{40\sqrt{m^{2}-q^{2}}}%
k\frac{q^{4}}{r_{\pm(0)}^{6}},
\end{equation}
where
\begin{equation}
r_{\pm(0)}=m\pm\sqrt{m^{2}-q^{2}}
\end{equation}
is the usual Reissner-Nordstr\"om horizon. Notice that
\begin{equation}
r_{+}>r_{+(0)}\ \text{\ e }\ r_{-}<r_{-(0)}.
\end{equation}
Hence, the effect of the nonlinearity of NLED is to displace the
external horizon outwards and the internal horizon inwards.

\subsection{Geometric redshift}

The geometric redshift of a light ray emitted radially at $r_{1}$
and detected at $r_{2}$ with $r_{+}<r_{1}<r_{2}$, is given by
\begin{equation}
1+z=\sqrt{\frac{\bar{g}_{00}\left( r_{2}\right) }{\bar{g}_{00}\left(
r_{1}\right) }}.\label{redshift}%
\end{equation}
By using Eq.~(\ref{g00}) and considering $r_{1}=r$ and
$r_{2}\rightarrow\infty$, one gets:
\begin{equation}
1+z\simeq\left[ 1-\frac{2m}{r}+\frac{q^{2}}{r^{2}}-\frac{k}{20}\frac{q^{4}%
}{r^{6}}-a_{\pm}\left( \frac{q}{r^{2}}\right) ^{2}W\left( r\right) \right] ^{-1/2}%
.\label{z_aprox1}%
\end{equation}
The first three terms of the right hand side of this equation
correspond to the usual Reissner-Nordstr\"om terms where the
redshift is increased by the mass while the charge (Maxwell term) is
responsible for decreasing it. The last two terms describe the
effects of the NLED: the fourth one is a consequence of the
curvature of spacetime generated by non-linear electromagnetic
terms and the fifth one is due to the effective metric of the NLED,
i.e., a redshift created by photons interacting with the background electric field. Note that both
terms lead to an increasing of $z$. It is also worth mentioning that the
influence of the effective metric makes redshift dependent on
the polarization since in general $a_{+}\neq a_{-}$.

If a weak field approximation is taken into account, i.e.
\begin{equation}
\frac{2m}{r}\ll1\text{ \ e \ }\frac{q^{2}}{r^{2}}\lll1,\label{campo fraco}%
\end{equation}
then Eq.~(\ref{z_aprox1}) is simplified to
\begin{equation}
z\simeq\frac{m}{r}+\frac{3}{2}\frac{m^{2}}{r^{2}}-\frac{q^{2}}{2r^{2}}%
+\frac{a_{\pm}}{2}\left( \frac{q}{r^{2}}\right) ^{2},\label{z weak field}
\end{equation}
where the mass dependence is taken up to second order and the term
$k\left( \frac{q}{r^{2}}\right) ^{2}\frac{q^{2}}{r^{2}}$ is
neglected, meaning that the curvature-born term disappears.

The magnitude of the last term in Eq.~(\ref{z weak field}) depends
on the coupling constant of the NLED, which is a free parameter (up
to constraints coming from experimental data). For instance, for
Born-Infeld Lagrangian, Eq.~(\ref{L_BI}),
\begin{equation}
a_{+}=a_{-}=\frac{1}{b^{2}}.
\end{equation}
The exception is Euler-Heisenberg electrodynamics which is obtained
as an effective theory at the low-energy regime of
QED.\footnote{Energies much less than that of the electron rest
energy.} According to
\citet{bialynicka1970nonlinear},
Euler-Heisenberg Lagrangian is given by
\begin{equation}
\mathcal{L}_{\text{EH}}\simeq F+\frac{2\alpha^{2}\hbar^{3}}{45m_{e}^{4}}\left(
4F^{2}+7G^{2}\right),  \label{L_EH}%
\end{equation}
where $\alpha$ is the fine-structure constant and $m_{e}$ is the
electron mass. This way,
\begin{equation}
a_{+}=\frac{16\alpha^{2}\hbar^{3}}{45m_{e}^{4}}\text{, \ }a_{-}=\frac
{28\alpha^{2}\hbar^{3}}{45m_{e}^{4}},%
\end{equation}
where $a_{+}\sim a_{-}\sim$
$10^{-28}\frac{\operatorname{cm}^{3}}{\operatorname{erg}}$.

In order to have an idea of the magnitude of the nonlinear term in
the context of Euler-Heisenberg electrodynamics, a hypothetical
situation is considered: a star like the Sun (with the same radius
$R_{\odot}$ and mass $M_{\odot}$) with an electric charge producing
an extremal black hole, i. e. $m=q$. This is the maximal charge
permitted for a static black hole if the cosmic censorship
hypothesis is considered. In this case,
\begin{align}
& \frac{m}{R_{\odot}}   \rightarrow \frac{M_{\odot}G}{c^{2}R_{\odot}}%
\simeq2\times10^{-6}  \Rightarrow \frac{m^{2}}{R_{\odot}^{2}}\sim\frac{q^{2}%
}{R_{\odot}^{2}}\sim10^{-12},  \\
& \text{ \ }\frac{a_{\pm}}{2}\left( \frac{q}{R_{\odot}^{2}}\right) ^{2}
\rightarrow\frac{a_{\pm}}{2}G\left( \frac{M_{\odot}}{R_{\odot}^{2}}\right)
^{2}\simeq4\times10^{-15}.
\end{align}
For this hypothetical situation, the nonlinear term (scaling with
$a_{\pm}$) is nine orders of magnitude smaller than the dominant
term $\frac{m}{r}$. Although being negligible this term becomes
rapidly important as $r$ decreases. In particular, for
$R_{\text{WD}}=10^{-3}R_{\odot}$ (a white dwarf radius) the nonlinear term
becomes of the same magnitude order of $\frac{m}{r}$, i.e.
\begin{align}
& \frac{m}{R_{\text{WD}}}   \sim \frac{a_{\pm}}{2}\left( \frac{q}{R_{\text{WD}}^{2}}\right) ^{2}\sim10^{-3}, \\
& \frac{m^{2}}{R_{\text{WD}}^{2}}  \sim \frac{q^{2}}{R_{\text{WD}}^{2}}\sim10^{-6}.
\end{align}
This might indicate an astrophysical object for which the NLED
effects could be important indeed.

Next, the geodesic deviation will be analyzed.

\subsection{Geodesic deviation of radiation}

The geodesic deviation of a light ray propagating from infinity to a
point at a distance $r$ from the origin of our coordinate system in
a region with static spherically symmetric gravitational and
electric fields is given by \citep{weinberg1972gravitation}:
\begin{equation}
\phi\left( r\right) -\phi_{\infty}=\int\limits_{r}^{\infty}\sqrt{\frac{\bar{g}%
_{11}\left( r\right) }{\left[ \left( \frac{r}{r_{0}}\right) ^{2}\left(
\frac{\bar{g}_{00}\left( r_{0}\right) }{\bar{g}_{00}\left( r\right) }\right)
-1\right] }}\frac{dr}{r}, \label{phi_r_bar}
\end{equation}
where $r_{0}$ is the maximum approximation distance of the light ray
from the gravitational/electric source. In the approximation of
small corrections to Maxwell theory $\bar{g}_{00}$ and
$\bar{g}_{11}$ are given by Eqs.~(\ref{g00}) and (\ref{-g11}),
respectively. In first order, the above expression is rewritten as
\begin{align}
& \left( \phi\left( r\right) -\phi_{\infty}\right) _{\pm} \simeq \nonumber \\
& \int \limits_{r}^{\infty}f_{1}\left( r\right) \left\{  1-a_{\pm}\left( \frac
{q}{r^{2}}\right) ^{2}f_{2}\left( r\right) +k\frac{q^{2}}{r^{2}}\left(
\frac{q}{r^{2}}\right) ^{2}f_{3}\left( r\right) \right\}  dr,\label{desvio1}%
\end{align}
where
\begin{align}
f_{1}\left( r\right)   &  =\frac{1}{r}\left[ \left( \frac{r}{r_{0}}\right)
^{2}W\left( r_{0}\right) -W\left( r\right) \right] ^{-1/2},\\
f_{2}\left( r\right)   &  =\frac{1}{2}+\frac{1}{2}\frac{W\left( r_{0}\right) -\left(
\frac{r}{r_{0}}\right) ^{4}W\left( r_{0}\right) }{W\left( r_{0}\right) -\left(
\frac{r_{0}}{r}\right) ^{2}W\left( r\right) },\\
f_{3}\left( r\right)   &  =\frac{1}{40W\left( r\right) }\left[ 1-
\frac{W\left( r_{0}\right) -\left( \frac{r}{r_{0}}\right) ^{6}W\left(
r\right) }{W\left( r_{0}\right) -\left( \frac{r_{0}}{r}\right) ^{2}W\left(
r\right) }\right] .
\end{align}
In Eq.~(\ref{desvio1}) the first term is the standard effect of
classical electrostatics in the context of general relativity, the
second term is the contribution from NLED effective metric, and
the third one is the correction due to spacetime curvature
coming from the nonlinear background electric field. Note that the
presence of the second term indicates that the geodesic deviation is
dependent on the polarization of the radiation.

The integral in Eq.~(\ref{desvio1}) can be evaluated analytically if
the weak field-approximation (\ref{campo fraco}) is considered:
\begin{align}
\left( \phi\left( r\right) -\phi_{\infty}\right) _{\pm} & \simeq
\arcsin\left(\frac{r_{0}}{r}\right) +\frac{m}{r_{0}}h_{1}\left( r\right) \nonumber \\
& -\frac{q^{2}}{2r_{0}^{2}}h_{2}\left( r\right) +\frac{1}{2}a_{\pm}\left( \frac{q}{r_{0}^{2}}\right) ^{2}h_{3}\left( r\right) ,\label{desvio 2}%
\end{align}
where
\begin{align}
h_{1}\left( r\right)  = & \, 2-\sqrt{1-\left( \frac{r_{0}}{r}\right) ^{2}}%
-\sqrt{\frac{1-\frac{r_{0}}{r}}{1+\frac{r_{0}}{r}}},\\
h_{2}\left( r\right)  = & \,\frac{3}{2}\arcsin\left(
\frac{r_{0}}{r}\right) -\frac{1}{2}\left( \frac{r_{0}}{r}\right)
\sqrt{1-\left( \frac{r_{0}}{r}\right) ^{2}} \\
h_{3}\left( r\right) = & \, \frac{9}{8}\arcsin\left( \frac{r_{0}}{r}\right)
-\frac{1}{8}\left( \frac{r_{0}}{r}\right) \sqrt{1-\left( \frac{r_{0}}%
{r}\right) ^{2}} \nonumber \\ & +\frac{1}{4}\sqrt{1-\left( \frac{r_{0}}{r}\right) ^{2}}\left(
\frac{r_{0}}{r}\right) ^{3}.
\end{align}

The deviation observed in a region far from the source is given by:
\begin{equation}
\Delta\phi=2\left\vert \phi\left( r_{0}\right) -\phi_{\infty}\right\vert
-\pi,
\end{equation}
resulting:
\begin{equation}
\Delta\phi_{_{\pm}}\simeq4\frac{m}{r_{0}}-\frac{3\pi}{4}\frac{q^{2}}{r_{0}^{2}}
 +a_{\pm}\frac{9\pi}{16}\left( \frac{q}{r_{0}^{2}%
}\right) ^{2}.\label{desvio 3}%
\end{equation}
This equation shows clearly the contributions of mass, electric
charge and the nonlinear term for the evaluation of geodesic
deviation in the weak-field
approximation. In the case of the Euler-Heisenberg Lagrangian, Eq.~(\ref{L_EH}%
), the contribution of each term can be evaluated by considerations
analogous to those of the previous section. Preliminary results of
the influence of the vacuum polarization effects described by
Euler-Heisenberg electrodynamics were obtained in
\citet{de2001dyadosphere}.

Particular cases are obtained if one takes one or more terms of
Eq.~(\ref{desvio 3}) to be null. For instance, for an object of null
charge the known result $\Delta\phi_{_{\pm}}=\frac{4m}{r_{0}}$ is
obtained. An interesting case is obtained when only the effect due
to NLED self-interaction (the interaction of the radiation with the
background electric field) is considered, i.e. no gravitational
effects are taken into account. Essentially, the metric considered
in this calculation is the effective metric in a flat spacetime.
Mathematically, this corresponds to take $W\left( r\right) =1$ and
$k=0$ in Eq.~(\ref{desvio1}). The integration of this expression
leads to a purely electric (PE) geodesic deviation:
\begin{equation}
\Delta\phi_{_{\pm}}^{\text{PE}}\simeq a_{\pm}\frac{9\pi}{16}\left( \frac{q}{r_{0}%
^{2}}\right) ^{2}.\label{desvio4}%
\end{equation}
Hypothetically, this result can be used to verify experimentally if
vacuum polarization is able to produce light deviation, as predicted
by Eq.~(\ref{desvio4}).

%\bigskip

\section{Conclusion}

In this paper we have constructed the solution of Einstein equations
for the exterior of spherically symmetric mass and charge
distributions in the context of non-linear electrodynamics. The
resulting line element is is given in Eqs.~(\ref{ds}), (\ref{nu})
and (\ref{S(r)}). It is asymptotically de Sitter and depends on the
mass $m$, on the cosmological constant $\Lambda$, and on the
functional $S(r)$. This object is an integral given in terms of the
electrostatic field $\vec{E}$ and the Lagrangian density
$\mathcal{L}(F,G^{2})$ of the particular NLED to be taken into
account.

First, we noticed that the radial null geodesics followed by photons
in the curved effective metric can be calculated as usual, being the
NLED influence encapsulated in function $\nu(r)$ via $S(r)$. The
non-linearity of electrodynamics appears explicitly in Eq.
(\ref{Sol_E}) for the magnitude of the electric field, which is
calculated when small perturbations of Maxwell theory are
considered. This leads to a modification in the horizons of the
charged BH as perceived by the photons: displacing the internal and
external horizons inwards and outwards respectively.

The redshift of spectral lines is increased by the presence of the
non-linear electrodynamic terms. Moreover, $z$ is sensible to the
polarization of the radiation due to self-interaction. In order to
estimate the magnitude of the effects on $z$ coming from NLED, we
decide to choose a particular non-linear theory and study it in the
weak-field regime.  Using Euler-Heisenberg Lagrangian, one
concludes that extremal compact charged objects could produce
redshifts for which the contribution from the ordinary $(m/r)$ term
is as important as the NLED-born term.

Finally, we calculated the geodesic deviation experienced by
radiation nearby massive charged bodies. The Reissner-Nordstr\"om
result is recovered along with small corrections to Maxwell
electrodynamics due to NLED. The background electric field affects
the geodesic path of light rays, increasing the deviation that would
be expected from the linear theory. We also obtained an expression
for the deviation when only the effect due to NLED self-interaction
is taken into account -- see Eq.~(\ref{desvio4}). This could be used
for comparison with observational data in order to determine if the
vacuum polarization actually can bend light rays.

\begin{acknowledgments}
KCV acknowledges CAPES-Brazil for financial support. LGM is grateful to FAPERN-Brazil for financial support.
\end{acknowledgments}

\bigskip

\bibliography{RefBHNLED}

\begin{thebibliography}{58}%
\makeatletter
\providecommand \@ifxundefined [1]{%
 \@ifx{#1\undefined}
}%
\providecommand \@ifnum [1]{%
 \ifnum #1\expandafter \@firstoftwo
 \else \expandafter \@secondoftwo
 \fi
}%
\providecommand \@ifx [1]{%
 \ifx #1\expandafter \@firstoftwo
 \else \expandafter \@secondoftwo
 \fi
}%
\providecommand \natexlab [1]{#1}%
\providecommand \enquote  [1]{``#1''}%
\providecommand \bibnamefont  [1]{#1}%
\providecommand \bibfnamefont [1]{#1}%
\providecommand \citenamefont [1]{#1}%
\providecommand \href@noop [0]{\@secondoftwo}%
\providecommand \href [0]{\begingroup \@sanitize@url \@href}%
\providecommand \@href[1]{\@@startlink{#1}\@@href}%
\providecommand \@@href[1]{\endgroup#1\@@endlink}%
\providecommand \@sanitize@url [0]{\catcode `\\12\catcode `\$12\catcode
  `\&12\catcode `\#12\catcode `\^12\catcode `\_12\catcode `\%12\relax}%
\providecommand \@@startlink[1]{}%
\providecommand \@@endlink[0]{}%
\providecommand \url  [0]{\begingroup\@sanitize@url \@url }%
\providecommand \@url [1]{\endgroup\@href {#1}{\urlprefix }}%
\providecommand \urlprefix  [0]{URL }%
\providecommand \Eprint [0]{\href }%
\providecommand \doibase [0]{http://dx.doi.org/}%
\providecommand \selectlanguage [0]{\@gobble}%
\providecommand \bibinfo  [0]{\@secondoftwo}%
\providecommand \bibfield  [0]{\@secondoftwo}%
\providecommand \translation [1]{[#1]}%
\providecommand \BibitemOpen [0]{}%
\providecommand \bibitemStop [0]{}%
\providecommand \bibitemNoStop [0]{.\EOS\space}%
\providecommand \EOS [0]{\spacefactor3000\relax}%
\providecommand \BibitemShut  [1]{\csname bibitem#1\endcsname}%
\let\auto@bib@innerbib\@empty
%</preamble>
\bibitem [{\citenamefont {Tu}\ \emph {et~al.}(2005)\citenamefont {Tu},
  \citenamefont {Luo},\ and\ \citenamefont {Gillies}}]{tu2005mass}%
  \BibitemOpen
  \bibfield  {author} {\bibinfo {author} {\bibfnamefont {L.-C.}\ \bibnamefont
  {Tu}}, \bibinfo {author} {\bibfnamefont {J.}~\bibnamefont {Luo}}, \ and\
  \bibinfo {author} {\bibfnamefont {G.~T.}\ \bibnamefont {Gillies}},\
  }\href@noop {} {\bibfield  {journal} {\bibinfo  {journal} {Reports on
  Progress in Physics}\ }\textbf {\bibinfo {volume} {68}},\ \bibinfo {pages}
  {77} (\bibinfo {year} {2005})}\BibitemShut {NoStop}%
\bibitem [{\citenamefont {Cuzinatto}\ \emph {et~al.}(2011)\citenamefont
  {Cuzinatto}, \citenamefont {Melo}, \citenamefont {Medeiros},\ and\
  \citenamefont {de~Pompeia}}]{cuzinatto2011probe}%
  \BibitemOpen
  \bibfield  {author} {\bibinfo {author} {\bibfnamefont {R.~R.}\ \bibnamefont
  {Cuzinatto}}, \bibinfo {author} {\bibfnamefont {C.~A.~M.}\ \bibnamefont
  {Melo}}, \bibinfo {author} {\bibfnamefont {L.~G.}\ \bibnamefont {Medeiros}},
  \ and\ \bibinfo {author} {\bibfnamefont {P.~J.}\ \bibnamefont {de~Pompeia}},\
  }\href@noop {} {\bibfield  {journal} {\bibinfo  {journal} {International
  Journal of Modern Physics A}\ }\textbf {\bibinfo {volume} {26}},\ \bibinfo
  {pages} {3641} (\bibinfo {year} {2011})}\BibitemShut {NoStop}%
\bibitem [{\citenamefont {Bonin}\ \emph {et~al.}(2010)\citenamefont {Bonin},
  \citenamefont {Bufalo}, \citenamefont {Pimentel},\ and\ \citenamefont
  {Zambrano}}]{bonin2010podolsky}%
  \BibitemOpen
  \bibfield  {author} {\bibinfo {author} {\bibfnamefont {C.~A.}\ \bibnamefont
  {Bonin}}, \bibinfo {author} {\bibfnamefont {R.}~\bibnamefont {Bufalo}},
  \bibinfo {author} {\bibfnamefont {B.~M.}\ \bibnamefont {Pimentel}}, \ and\
  \bibinfo {author} {\bibfnamefont {G.~E.~R.}\ \bibnamefont {Zambrano}},\
  }\href@noop {} {\bibfield  {journal} {\bibinfo  {journal} {Physical Review
  D}\ }\textbf {\bibinfo {volume} {81}},\ \bibinfo {pages} {025003} (\bibinfo
  {year} {2010})}\BibitemShut {NoStop}%
\bibitem [{\citenamefont {Heisenberg}\ and\ \citenamefont
  {Euler}(1936)}]{heisenberg1936folgerungen}%
  \BibitemOpen
  \bibfield  {author} {\bibinfo {author} {\bibfnamefont {W.}~\bibnamefont
  {Heisenberg}}\ and\ \bibinfo {author} {\bibfnamefont {H.}~\bibnamefont
  {Euler}},\ }\href@noop {} {\bibfield  {journal} {\bibinfo  {journal}
  {Zeitschrift f{\"u}r Physik}\ }\textbf {\bibinfo {volume} {98}},\ \bibinfo
  {pages} {714} (\bibinfo {year} {1936})}\BibitemShut {NoStop}%
\bibitem [{\citenamefont {Schwinger}(1951)}]{schwinger1951}%
  \BibitemOpen
  \bibfield  {author} {\bibinfo {author} {\bibfnamefont {J.}~\bibnamefont
  {Schwinger}},\ }\href@noop {} {\bibfield  {journal} {\bibinfo  {journal}
  {Physical Review}\ }\textbf {\bibinfo {volume} {82}},\ \bibinfo {pages} {664}
  (\bibinfo {year} {1951})}\BibitemShut {NoStop}%
\bibitem [{\citenamefont {Seiberg}\ and\ \citenamefont
  {Witten}(1999)}]{seiberg1999stringnoncom}%
  \BibitemOpen
  \bibfield  {author} {\bibinfo {author} {\bibfnamefont {N.}~\bibnamefont
  {Seiberg}}\ and\ \bibinfo {author} {\bibfnamefont {E.}~\bibnamefont
  {Witten}},\ }\href@noop {} {\bibfield  {journal} {\bibinfo  {journal}
  {Journal of High Energy Physics}\ }\textbf {\bibinfo {volume} {1999}},\
  \bibinfo {pages} {032} (\bibinfo {year} {1999})}\BibitemShut {NoStop}%
\bibitem [{\citenamefont {Fradkin}\ and\ \citenamefont
  {Tseytlin}(1985)}]{fradkin1985non}%
  \BibitemOpen
  \bibfield  {author} {\bibinfo {author} {\bibfnamefont {E.~S.}\ \bibnamefont
  {Fradkin}}\ and\ \bibinfo {author} {\bibfnamefont {A.~A.}\ \bibnamefont
  {Tseytlin}},\ }\href@noop {} {\bibfield  {journal} {\bibinfo  {journal}
  {Physics Letters B}\ }\textbf {\bibinfo {volume} {163}},\ \bibinfo {pages}
  {123} (\bibinfo {year} {1985})}\BibitemShut {NoStop}%
\bibitem [{\citenamefont {Bergshoeff}\ \emph {et~al.}(1987)\citenamefont
  {Bergshoeff}, \citenamefont {Sezgin}, \citenamefont {Pope},\ and\
  \citenamefont {Townsend}}]{bergshoeff1987born}%
  \BibitemOpen
  \bibfield  {author} {\bibinfo {author} {\bibfnamefont {E.}~\bibnamefont
  {Bergshoeff}}, \bibinfo {author} {\bibfnamefont {E.}~\bibnamefont {Sezgin}},
  \bibinfo {author} {\bibfnamefont {C.~N.}\ \bibnamefont {Pope}}, \ and\
  \bibinfo {author} {\bibfnamefont {P.~K.}\ \bibnamefont {Townsend}},\
  }\href@noop {} {\bibfield  {journal} {\bibinfo  {journal} {Physics Letters
  B}\ }\textbf {\bibinfo {volume} {188}},\ \bibinfo {pages} {70} (\bibinfo
  {year} {1987})}\BibitemShut {NoStop}%
\bibitem [{\citenamefont {Metsaev}\ \emph {et~al.}(1987)\citenamefont
  {Metsaev}, \citenamefont {Rahmanov},\ and\ \citenamefont
  {Tseytlin}}]{metsaev1987born}%
  \BibitemOpen
  \bibfield  {author} {\bibinfo {author} {\bibfnamefont {R.~R.}\ \bibnamefont
  {Metsaev}}, \bibinfo {author} {\bibfnamefont {M.~A.}\ \bibnamefont
  {Rahmanov}}, \ and\ \bibinfo {author} {\bibfnamefont {A.~A.}\ \bibnamefont
  {Tseytlin}},\ }\href@noop {} {\bibfield  {journal} {\bibinfo  {journal}
  {Physics Letters B}\ }\textbf {\bibinfo {volume} {193}},\ \bibinfo {pages}
  {207} (\bibinfo {year} {1987})}\BibitemShut {NoStop}%
\bibitem [{\citenamefont {Leigh}(1989)}]{leigh1989dirac}%
  \BibitemOpen
  \bibfield  {author} {\bibinfo {author} {\bibfnamefont {R.~G.}\ \bibnamefont
  {Leigh}},\ }\href@noop {} {\bibfield  {journal} {\bibinfo  {journal} {Modern
  Physics Letters A}\ }\textbf {\bibinfo {volume} {4}},\ \bibinfo {pages}
  {2767} (\bibinfo {year} {1989})}\BibitemShut {NoStop}%
\bibitem [{\citenamefont {Born}(1934)}]{born1934royal}%
  \BibitemOpen
  \bibfield  {author} {\bibinfo {author} {\bibfnamefont {M.}~\bibnamefont
  {Born}},\ }\href@noop {} {\bibfield  {journal} {\bibinfo  {journal}
  {Proceedings of the Royal Society of London. Series A}\ }\textbf {\bibinfo
  {volume} {143}},\ \bibinfo {pages} {410} (\bibinfo {year}
  {1934})}\BibitemShut {NoStop}%
\bibitem [{\citenamefont {Born}(1937)}]{born1939poincare}%
  \BibitemOpen
  \bibfield  {author} {\bibinfo {author} {\bibfnamefont {M.}~\bibnamefont
  {Born}},\ }\href@noop {} {\bibfield  {journal} {\bibinfo  {journal} {Annales
  de l'Institut Henri Poincar\'e}\ }\textbf {\bibinfo {volume} {7}},\ \bibinfo
  {pages} {155} (\bibinfo {year} {1937})}\BibitemShut {NoStop}%
\bibitem [{\citenamefont {Born}\ and\ \citenamefont
  {Infeld}(1934)}]{born1934foundations}%
  \BibitemOpen
  \bibfield  {author} {\bibinfo {author} {\bibfnamefont {M.}~\bibnamefont
  {Born}}\ and\ \bibinfo {author} {\bibfnamefont {L.}~\bibnamefont {Infeld}},\
  }\href@noop {} {\bibfield  {journal} {\bibinfo  {journal} {Proceedings of the
  Royal Society of London. Series A}\ }\textbf {\bibinfo {volume} {144}},\
  \bibinfo {pages} {425} (\bibinfo {year} {1934})}\BibitemShut {NoStop}%
\bibitem [{\citenamefont {Stehle}\ and\ \citenamefont
  {DeBaryshe}(1966)}]{Stehle1966QEDCorrespondence}%
  \BibitemOpen
  \bibfield  {author} {\bibinfo {author} {\bibfnamefont {P.}~\bibnamefont
  {Stehle}}\ and\ \bibinfo {author} {\bibfnamefont {P.~G.}\ \bibnamefont
  {DeBaryshe}},\ }\href@noop {} {\bibfield  {journal} {\bibinfo  {journal}
  {Phys. Rev.}\ }\textbf {\bibinfo {volume} {152}},\ \bibinfo {pages} {1135}
  (\bibinfo {year} {1966})}\BibitemShut {NoStop}%
\bibitem [{\citenamefont {Delphenich}(2003)}]{delphenich2003nonqed}%
  \BibitemOpen
  \bibfield  {author} {\bibinfo {author} {\bibfnamefont {D.~H.}\ \bibnamefont
  {Delphenich}},\ }\href@noop {} {\bibfield  {journal} {\bibinfo  {journal}
  {arXiv preprint hep-th/0309108}\ } (\bibinfo {year} {2003})}\BibitemShut
  {NoStop}%
\bibitem [{\citenamefont {Dunne}(2005)}]{dunne2005heisenberg}%
  \BibitemOpen
  \bibfield  {author} {\bibinfo {author} {\bibfnamefont {G.~V.}\ \bibnamefont
  {Dunne}},\ }\href@noop {} {\bibfield  {journal} {\bibinfo  {journal} {From
  fields to strings: Circumnavigating theoretical physics}\ }\textbf {\bibinfo
  {volume} {1}},\ \bibinfo {pages} {445} (\bibinfo {year} {2005})}\BibitemShut
  {NoStop}%
\bibitem [{\citenamefont {Salazar}\ \emph {et~al.}(1984)\citenamefont
  {Salazar}, \citenamefont {Garc{\'\i}a},\ and\ \citenamefont
  {Pleba{\'n}ski}}]{plebanski1984type}%
  \BibitemOpen
  \bibfield  {author} {\bibinfo {author} {\bibfnamefont {H.}~\bibnamefont
  {Salazar}}, \bibinfo {author} {\bibfnamefont {A.}~\bibnamefont
  {Garc{\'\i}a}}, \ and\ \bibinfo {author} {\bibfnamefont {J.}~\bibnamefont
  {Pleba{\'n}ski}},\ }\href@noop {} {\bibfield  {journal} {\bibinfo  {journal}
  {Il Nuovo Cimento B Series 11}\ }\textbf {\bibinfo {volume} {84}},\ \bibinfo
  {pages} {65} (\bibinfo {year} {1984})}\BibitemShut {NoStop}%
\bibitem [{\citenamefont {Demianski}(1986)}]{demianski1986static}%
  \BibitemOpen
  \bibfield  {author} {\bibinfo {author} {\bibfnamefont {M.}~\bibnamefont
  {Demianski}},\ }\href@noop {} {\bibfield  {journal} {\bibinfo  {journal}
  {Foundations of physics}\ }\textbf {\bibinfo {volume} {16}},\ \bibinfo
  {pages} {187} (\bibinfo {year} {1986})}\BibitemShut {NoStop}%
\bibitem [{\citenamefont {De~Oliveira}(1994)}]{oliveira1994non}%
  \BibitemOpen
  \bibfield  {author} {\bibinfo {author} {\bibfnamefont {H.~P.}\ \bibnamefont
  {De~Oliveira}},\ }\href@noop {} {\bibfield  {journal} {\bibinfo  {journal}
  {Classical and Quantum Gravity}\ }\textbf {\bibinfo {volume} {11}},\ \bibinfo
  {pages} {1469} (\bibinfo {year} {1994})}\BibitemShut {NoStop}%
\bibitem [{\citenamefont {Novello}\ \emph {et~al.}(2003)\citenamefont
  {Novello}, \citenamefont {Perez~Bergliaffa}, \citenamefont {Salim},
  \citenamefont {De~Lorenci},\ and\ \citenamefont
  {Klippert}}]{novello2002analogbh}%
  \BibitemOpen
  \bibfield  {author} {\bibinfo {author} {\bibfnamefont {M.}~\bibnamefont
  {Novello}}, \bibinfo {author} {\bibfnamefont {S.~E.}\ \bibnamefont
  {Perez~Bergliaffa}}, \bibinfo {author} {\bibfnamefont {J.}~\bibnamefont
  {Salim}}, \bibinfo {author} {\bibfnamefont {V.}~\bibnamefont {De~Lorenci}}, \
  and\ \bibinfo {author} {\bibfnamefont {R.}~\bibnamefont {Klippert}},\
  }\href@noop {} {\bibfield  {journal} {\bibinfo  {journal} {Class. Quant.
  Grav.}\ }\textbf {\bibinfo {volume} {20}},\ \bibinfo {pages} {859} (\bibinfo
  {year} {2003})}\BibitemShut {NoStop}%
\bibitem [{\citenamefont {Barcelo}\ \emph {et~al.}(2005)\citenamefont
  {Barcelo}, \citenamefont {Liberati},\ and\ \citenamefont
  {Visser}}]{barcelo2005analoggravity}%
  \BibitemOpen
  \bibfield  {author} {\bibinfo {author} {\bibfnamefont {C.}~\bibnamefont
  {Barcelo}}, \bibinfo {author} {\bibfnamefont {S.}~\bibnamefont {Liberati}}, \
  and\ \bibinfo {author} {\bibfnamefont {M.}~\bibnamefont {Visser}},\
  }\href@noop {} {\bibfield  {journal} {\bibinfo  {journal} {Living Rev. Rel.}\
  }\textbf {\bibinfo {volume} {8}},\ \bibinfo {pages} {12} (\bibinfo {year}
  {2005})}\BibitemShut {NoStop}%
\bibitem [{\citenamefont {Cai}\ \emph {et~al.}(2008)\citenamefont {Cai},
  \citenamefont {Nie},\ and\ \citenamefont {Sun}}]{cai2008shearviscosity}%
  \BibitemOpen
  \bibfield  {author} {\bibinfo {author} {\bibfnamefont {R.-G.}\ \bibnamefont
  {Cai}}, \bibinfo {author} {\bibfnamefont {Z.-Y.}\ \bibnamefont {Nie}}, \ and\
  \bibinfo {author} {\bibfnamefont {Y.-W.}\ \bibnamefont {Sun}},\ }\href@noop
  {} {\bibfield  {journal} {\bibinfo  {journal} {Phys. Rev. D}\ }\textbf
  {\bibinfo {volume} {78}},\ \bibinfo {pages} {126007} (\bibinfo {year}
  {2008})}\BibitemShut {NoStop}%
\bibitem [{\citenamefont {Garcia-Salcedo}\ and\ \citenamefont
  {Breton}(2000)}]{garcia2000born}%
  \BibitemOpen
  \bibfield  {author} {\bibinfo {author} {\bibfnamefont {R.}~\bibnamefont
  {Garcia-Salcedo}}\ and\ \bibinfo {author} {\bibfnamefont {N.}~\bibnamefont
  {Breton}},\ }\href@noop {} {\bibfield  {journal} {\bibinfo  {journal}
  {International Journal of Modern Physics A}\ }\textbf {\bibinfo {volume}
  {15}},\ \bibinfo {pages} {4341} (\bibinfo {year} {2000})}\BibitemShut
  {NoStop}%
\bibitem [{\citenamefont {De~Lorenci}\ \emph {et~al.}(2002)\citenamefont
  {De~Lorenci}, \citenamefont {Klippert}, \citenamefont {Novello},\ and\
  \citenamefont {Salim}}]{delorenci2002nonlinear}%
  \BibitemOpen
  \bibfield  {author} {\bibinfo {author} {\bibfnamefont {V.~A.}\ \bibnamefont
  {De~Lorenci}}, \bibinfo {author} {\bibfnamefont {R.}~\bibnamefont
  {Klippert}}, \bibinfo {author} {\bibfnamefont {M.}~\bibnamefont {Novello}}, \
  and\ \bibinfo {author} {\bibfnamefont {J.~M.}\ \bibnamefont {Salim}},\
  }\href@noop {} {\bibfield  {journal} {\bibinfo  {journal} {Physical Review
  D}\ }\textbf {\bibinfo {volume} {65}},\ \bibinfo {pages} {063501} (\bibinfo
  {year} {2002})}\BibitemShut {NoStop}%
\bibitem [{\citenamefont {Dyadichev}\ \emph {et~al.}(2002)\citenamefont
  {Dyadichev}, \citenamefont {Galtsov}, \citenamefont {Zorin},\ and\
  \citenamefont {Zotov}}]{dyadichev2002non}%
  \BibitemOpen
  \bibfield  {author} {\bibinfo {author} {\bibfnamefont {V.~V.}\ \bibnamefont
  {Dyadichev}}, \bibinfo {author} {\bibfnamefont {D.~V.}\ \bibnamefont
  {Galtsov}}, \bibinfo {author} {\bibfnamefont {A.~G.}\ \bibnamefont {Zorin}},
  \ and\ \bibinfo {author} {\bibfnamefont {M.~Y.}\ \bibnamefont {Zotov}},\
  }\href@noop {} {\bibfield  {journal} {\bibinfo  {journal} {Physical Review
  D}\ }\textbf {\bibinfo {volume} {65}},\ \bibinfo {pages} {084007} (\bibinfo
  {year} {2002})}\BibitemShut {NoStop}%
\bibitem [{\citenamefont {Moniz}(2002)}]{moniz2002quintessence}%
  \BibitemOpen
  \bibfield  {author} {\bibinfo {author} {\bibfnamefont {P.~V.}\ \bibnamefont
  {Moniz}},\ }\href@noop {} {\bibfield  {journal} {\bibinfo  {journal}
  {Physical Review D}\ }\textbf {\bibinfo {volume} {66}},\ \bibinfo {pages}
  {103501} (\bibinfo {year} {2002})}\BibitemShut {NoStop}%
\bibitem [{\citenamefont {Novello}\ \emph {et~al.}(2004)\citenamefont
  {Novello}, \citenamefont {Bergliaffa},\ and\ \citenamefont
  {Salim}}]{novello2004nonlinear}%
  \BibitemOpen
  \bibfield  {author} {\bibinfo {author} {\bibfnamefont {M.}~\bibnamefont
  {Novello}}, \bibinfo {author} {\bibfnamefont {S.~E.~P.}\ \bibnamefont
  {Bergliaffa}}, \ and\ \bibinfo {author} {\bibfnamefont {J.}~\bibnamefont
  {Salim}},\ }\href@noop {} {\bibfield  {journal} {\bibinfo  {journal}
  {Physical Review D}\ }\textbf {\bibinfo {volume} {69}},\ \bibinfo {pages}
  {127301} (\bibinfo {year} {2004})}\BibitemShut {NoStop}%
\bibitem [{\citenamefont {Novello}\ \emph {et~al.}(2009)\citenamefont
  {Novello}, \citenamefont {Araujo},\ and\ \citenamefont
  {Salim}}]{novello2009cyclic}%
  \BibitemOpen
  \bibfield  {author} {\bibinfo {author} {\bibfnamefont {M.}~\bibnamefont
  {Novello}}, \bibinfo {author} {\bibfnamefont {A.~N.}\ \bibnamefont {Araujo}},
  \ and\ \bibinfo {author} {\bibfnamefont {J.~M.}\ \bibnamefont {Salim}},\
  }\href@noop {} {\bibfield  {journal} {\bibinfo  {journal} {International
  Journal of Modern Physics A}\ }\textbf {\bibinfo {volume} {24}},\ \bibinfo
  {pages} {5639} (\bibinfo {year} {2009})}\BibitemShut {NoStop}%
\bibitem [{\citenamefont {Medeiros}(2012)}]{medeiros2012realistic}%
  \BibitemOpen
  \bibfield  {author} {\bibinfo {author} {\bibfnamefont {L.~G.}\ \bibnamefont
  {Medeiros}},\ }\href@noop {} {\bibfield  {journal} {\bibinfo  {journal}
  {International Journal of Modern Physics D}\ }\textbf {\bibinfo {volume}
  {21}} (\bibinfo {year} {2012})}\BibitemShut {NoStop}%
\bibitem [{\citenamefont {Ayon-Beato}\ and\ \citenamefont
  {Garcia}(1998)}]{ayon1998regular}%
  \BibitemOpen
  \bibfield  {author} {\bibinfo {author} {\bibfnamefont {E.}~\bibnamefont
  {Ayon-Beato}}\ and\ \bibinfo {author} {\bibfnamefont {A.}~\bibnamefont
  {Garcia}},\ }\href@noop {} {\bibfield  {journal} {\bibinfo  {journal}
  {Physical Review Letters}\ }\textbf {\bibinfo {volume} {80}},\ \bibinfo
  {pages} {5056} (\bibinfo {year} {1998})}\BibitemShut {NoStop}%
\bibitem [{\citenamefont {Yajima}\ and\ \citenamefont
  {Tamaki}(2001)}]{yajima2001black}%
  \BibitemOpen
  \bibfield  {author} {\bibinfo {author} {\bibfnamefont {H.}~\bibnamefont
  {Yajima}}\ and\ \bibinfo {author} {\bibfnamefont {T.}~\bibnamefont
  {Tamaki}},\ }\href@noop {} {\bibfield  {journal} {\bibinfo  {journal}
  {Physical Review D}\ }\textbf {\bibinfo {volume} {63}},\ \bibinfo {pages}
  {064007} (\bibinfo {year} {2001})}\BibitemShut {NoStop}%
\bibitem [{\citenamefont {Bronnikov}(2001)}]{bronnikov2001regular}%
  \BibitemOpen
  \bibfield  {author} {\bibinfo {author} {\bibfnamefont {K.~A.}\ \bibnamefont
  {Bronnikov}},\ }\href@noop {} {\bibfield  {journal} {\bibinfo  {journal}
  {Physical Review D}\ }\textbf {\bibinfo {volume} {63}},\ \bibinfo {pages}
  {044005} (\bibinfo {year} {2001})}\BibitemShut {NoStop}%
\bibitem [{\citenamefont {Diaz-Alonso}\ and\ \citenamefont
  {Rubiera-Garcia}(2010{\natexlab{a}})}]{diaz2010electrostatic}%
  \BibitemOpen
  \bibfield  {author} {\bibinfo {author} {\bibfnamefont {J.}~\bibnamefont
  {Diaz-Alonso}}\ and\ \bibinfo {author} {\bibfnamefont {D.}~\bibnamefont
  {Rubiera-Garcia}},\ }\href@noop {} {\bibfield  {journal} {\bibinfo  {journal}
  {Physical Review D}\ }\textbf {\bibinfo {volume} {81}},\ \bibinfo {pages}
  {064021} (\bibinfo {year} {2010}{\natexlab{a}})}\BibitemShut {NoStop}%
\bibitem [{\citenamefont {Diaz-Alonso}\ and\ \citenamefont
  {Rubiera-Garcia}(2010{\natexlab{b}})}]{diaz2010asymptotically}%
  \BibitemOpen
  \bibfield  {author} {\bibinfo {author} {\bibfnamefont {J.}~\bibnamefont
  {Diaz-Alonso}}\ and\ \bibinfo {author} {\bibfnamefont {D.}~\bibnamefont
  {Rubiera-Garcia}},\ }\href@noop {} {\bibfield  {journal} {\bibinfo  {journal}
  {Physical Review D}\ }\textbf {\bibinfo {volume} {82}},\ \bibinfo {pages}
  {085024} (\bibinfo {year} {2010}{\natexlab{b}})}\BibitemShut {NoStop}%
\bibitem [{\citenamefont {Diaz-Alonso}\ and\ \citenamefont
  {Rubiera-Garcia}(2013{\natexlab{a}})}]{diaz2013charged}%
  \BibitemOpen
  \bibfield  {author} {\bibinfo {author} {\bibfnamefont {J.}~\bibnamefont
  {Diaz-Alonso}}\ and\ \bibinfo {author} {\bibfnamefont {D.}~\bibnamefont
  {Rubiera-Garcia}},\ }\href@noop {} {\bibfield  {journal} {\bibinfo  {journal}
  {Journal of Physics: Conference Series}\ }\textbf {\bibinfo {volume} {314}},\
  \bibinfo {pages} {012065} (\bibinfo {year} {2013}{\natexlab{a}})}\BibitemShut
  {NoStop}%
\bibitem [{\citenamefont {Ruffini}\ \emph {et~al.}(2013)\citenamefont
  {Ruffini}, \citenamefont {Wu},\ and\ \citenamefont
  {Xue}}]{ruffini2013einstein}%
  \BibitemOpen
  \bibfield  {author} {\bibinfo {author} {\bibfnamefont {R.}~\bibnamefont
  {Ruffini}}, \bibinfo {author} {\bibfnamefont {Y.-B.}\ \bibnamefont {Wu}}, \
  and\ \bibinfo {author} {\bibfnamefont {S.-S.}\ \bibnamefont {Xue}},\
  }\href@noop {} {\bibfield  {journal} {\bibinfo  {journal} {Physical Review
  D}\ }\textbf {\bibinfo {volume} {88}},\ \bibinfo {pages} {085004} (\bibinfo
  {year} {2013})}\BibitemShut {NoStop}%
\bibitem [{\citenamefont {Chemissany}\ \emph {et~al.}(2008)\citenamefont
  {Chemissany}, \citenamefont {De~Roo},\ and\ \citenamefont
  {Panda}}]{chemissany2008thermodynamics}%
  \BibitemOpen
  \bibfield  {author} {\bibinfo {author} {\bibfnamefont {W.~A.}\ \bibnamefont
  {Chemissany}}, \bibinfo {author} {\bibfnamefont {M.}~\bibnamefont {De~Roo}},
  \ and\ \bibinfo {author} {\bibfnamefont {S.}~\bibnamefont {Panda}},\
  }\href@noop {} {\bibfield  {journal} {\bibinfo  {journal} {Classical and
  Quantum Gravity}\ }\textbf {\bibinfo {volume} {25}},\ \bibinfo {pages}
  {225009} (\bibinfo {year} {2008})}\BibitemShut {NoStop}%
\bibitem [{\citenamefont {Gunasekaran}\ \emph {et~al.}(2012)\citenamefont
  {Gunasekaran}, \citenamefont {Kubiz{\v{n}}{\'a}k},\ and\ \citenamefont
  {Mann}}]{gunasekaran2012extended}%
  \BibitemOpen
  \bibfield  {author} {\bibinfo {author} {\bibfnamefont {S.}~\bibnamefont
  {Gunasekaran}}, \bibinfo {author} {\bibfnamefont {D.}~\bibnamefont
  {Kubiz{\v{n}}{\'a}k}}, \ and\ \bibinfo {author} {\bibfnamefont {R.~B.}\
  \bibnamefont {Mann}},\ }\href@noop {} {\bibfield  {journal} {\bibinfo
  {journal} {Journal of High Energy Physics}\ }\textbf {\bibinfo {volume}
  {1211}},\ \bibinfo {pages} {110} (\bibinfo {year} {2012})}\BibitemShut
  {NoStop}%
\bibitem [{\citenamefont {Hendi}\ and\ \citenamefont
  {Vahidinia}(2013)}]{Hendi2012}%
  \BibitemOpen
  \bibfield  {author} {\bibinfo {author} {\bibfnamefont {S.~H.}\ \bibnamefont
  {Hendi}}\ and\ \bibinfo {author} {\bibfnamefont {M.~H.}\ \bibnamefont
  {Vahidinia}},\ }\href@noop {} {\bibfield  {journal} {\bibinfo  {journal}
  {Physical Review D}\ }\textbf {\bibinfo {volume} {88}},\ \bibinfo {pages}
  {84045} (\bibinfo {year} {2013})}\BibitemShut {NoStop}%
\bibitem [{\citenamefont {Diaz-Alonso}\ and\ \citenamefont
  {Rubiera-Garcia}(2013{\natexlab{b}})}]{diaz2013thermodynamic}%
  \BibitemOpen
  \bibfield  {author} {\bibinfo {author} {\bibfnamefont {J.}~\bibnamefont
  {Diaz-Alonso}}\ and\ \bibinfo {author} {\bibfnamefont {D.}~\bibnamefont
  {Rubiera-Garcia}},\ }\href@noop {} {\bibfield  {journal} {\bibinfo  {journal}
  {General Relativity and Gravitation}\ }\textbf {\bibinfo {volume} {45}},\
  \bibinfo {pages} {1901} (\bibinfo {year} {2013}{\natexlab{b}})}\BibitemShut
  {NoStop}%
\bibitem [{\citenamefont {Mo}\ and\ \citenamefont {Liu}(2014)}]{Jie2014}%
  \BibitemOpen
  \bibfield  {author} {\bibinfo {author} {\bibfnamefont {J.~X.}\ \bibnamefont
  {Mo}}\ and\ \bibinfo {author} {\bibfnamefont {W.~B.}\ \bibnamefont {Liu}},\
  }\href@noop {} {\bibfield  {journal} {\bibinfo  {journal} {The European
  Physical Journal C}\ }\textbf {\bibinfo {volume} {74}},\ \bibinfo {pages}
  {2836} (\bibinfo {year} {2014})}\BibitemShut {NoStop}%
\bibitem [{\citenamefont {Bret{\'o}n}\ and\ \citenamefont
  {Bergliaffa}(2014)}]{breton2014stability}%
  \BibitemOpen
  \bibfield  {author} {\bibinfo {author} {\bibfnamefont {N.}~\bibnamefont
  {Bret{\'o}n}}\ and\ \bibinfo {author} {\bibfnamefont {S.~E.~P.}\ \bibnamefont
  {Bergliaffa}},\ }\href@noop {} {\bibfield  {journal} {\bibinfo  {journal}
  {preprint}\ } (\bibinfo {year} {2014})}\BibitemShut {NoStop}%
\bibitem [{\citenamefont {Hendi}\ \emph {et~al.}(2014)\citenamefont {Hendi},
  \citenamefont {Panahiyan},\ and\ \citenamefont {Mahmoudi}}]{Hendi2014}%
  \BibitemOpen
  \bibfield  {author} {\bibinfo {author} {\bibfnamefont {S.~H.}\ \bibnamefont
  {Hendi}}, \bibinfo {author} {\bibfnamefont {S.}~\bibnamefont {Panahiyan}}, \
  and\ \bibinfo {author} {\bibfnamefont {E.}~\bibnamefont {Mahmoudi}},\
  }\href@noop {} {\bibfield  {journal} {\bibinfo  {journal} {The European
  Physical Journal C}\ }\textbf {\bibinfo {volume} {74}},\ \bibinfo {pages}
  {3079} (\bibinfo {year} {2014})}\BibitemShut {NoStop}%
\bibitem [{\citenamefont {Hendi}\ and\ \citenamefont
  {Momennia}(2015)}]{Hendi2015}%
  \BibitemOpen
  \bibfield  {author} {\bibinfo {author} {\bibfnamefont {S.~H.}\ \bibnamefont
  {Hendi}}\ and\ \bibinfo {author} {\bibfnamefont {M.}~\bibnamefont
  {Momennia}},\ }\href@noop {} {\bibfield  {journal} {\bibinfo  {journal} {The
  European Physical Journal C}\ }\textbf {\bibinfo {volume} {75}},\ \bibinfo
  {pages} {54} (\bibinfo {year} {2015})}\BibitemShut {NoStop}%
\bibitem [{\citenamefont {Hendi}\ and\ \citenamefont
  {Panahiyan}(2014)}]{Hendi2015a}%
  \BibitemOpen
  \bibfield  {author} {\bibinfo {author} {\bibfnamefont {S.~H.}\ \bibnamefont
  {Hendi}}\ and\ \bibinfo {author} {\bibfnamefont {S.}~\bibnamefont
  {Panahiyan}},\ }\href@noop {} {\bibfield  {journal} {\bibinfo  {journal}
  {Physical Review D}\ }\textbf {\bibinfo {volume} {90}},\ \bibinfo {pages}
  {124008} (\bibinfo {year} {2014})}\BibitemShut {NoStop}%
\bibitem [{\citenamefont {Linares}\ \emph {et~al.}(2014)\citenamefont
  {Linares}, \citenamefont {Maceda},\ and\ \citenamefont
  {Mart{\'\i}nez-Carbajal}}]{linares2014test}%
  \BibitemOpen
  \bibfield  {author} {\bibinfo {author} {\bibfnamefont {R.}~\bibnamefont
  {Linares}}, \bibinfo {author} {\bibfnamefont {M.}~\bibnamefont {Maceda}}, \
  and\ \bibinfo {author} {\bibfnamefont {D.}~\bibnamefont
  {Mart{\'\i}nez-Carbajal}},\ }\href@noop {} {\bibfield  {journal} {\bibinfo
  {journal} {preprint}\ } (\bibinfo {year} {2014})}\BibitemShut {NoStop}%
\bibitem [{\citenamefont {Bialynicka~Birula}\ and\ \citenamefont
  {Bialynicki~Birula}(1970)}]{bialynicka1970nonlinear}%
  \BibitemOpen
  \bibfield  {author} {\bibinfo {author} {\bibfnamefont {Z.}~\bibnamefont
  {Bialynicka~Birula}}\ and\ \bibinfo {author} {\bibfnamefont {I.}~\bibnamefont
  {Bialynicki~Birula}},\ }\href@noop {} {\bibfield  {journal} {\bibinfo
  {journal} {Physical Review D}\ }\textbf {\bibinfo {volume} {2}},\ \bibinfo
  {pages} {2341} (\bibinfo {year} {1970})}\BibitemShut {NoStop}%
\bibitem [{\citenamefont {Boillat}(1970)}]{boillat1970nonlinear}%
  \BibitemOpen
  \bibfield  {author} {\bibinfo {author} {\bibfnamefont {G.}~\bibnamefont
  {Boillat}},\ }\href@noop {} {\bibfield  {journal} {\bibinfo  {journal}
  {Journal of Mathematical Physics}\ }\textbf {\bibinfo {volume} {11}},\
  \bibinfo {pages} {941} (\bibinfo {year} {1970})}\BibitemShut {NoStop}%
\bibitem [{\citenamefont {Novello}\ \emph {et~al.}(2000)\citenamefont
  {Novello}, \citenamefont {De~Lorenci}, \citenamefont {Salim},\ and\
  \citenamefont {Klippert}}]{novello2000geometrical}%
  \BibitemOpen
  \bibfield  {author} {\bibinfo {author} {\bibfnamefont {M.}~\bibnamefont
  {Novello}}, \bibinfo {author} {\bibfnamefont {V.~A.}\ \bibnamefont
  {De~Lorenci}}, \bibinfo {author} {\bibfnamefont {J.~M.}\ \bibnamefont
  {Salim}}, \ and\ \bibinfo {author} {\bibfnamefont {R.}~\bibnamefont
  {Klippert}},\ }\href@noop {} {\bibfield  {journal} {\bibinfo  {journal}
  {Physical Review D}\ }\textbf {\bibinfo {volume} {61}},\ \bibinfo {pages}
  {045001} (\bibinfo {year} {2000})}\BibitemShut {NoStop}%
\bibitem [{\citenamefont {Novello}\ and\ \citenamefont
  {Goulart}(2010)}]{novello2010book}%
  \BibitemOpen
  \bibfield  {author} {\bibinfo {author} {\bibfnamefont {M.}~\bibnamefont
  {Novello}}\ and\ \bibinfo {author} {\bibfnamefont {E.}~\bibnamefont
  {Goulart}},\ }\href@noop {} {\emph {\bibinfo {title} {Eletrodin\^amica n\~ao
  linear. Causalidade e efeitos cosmol\'ogicos}}}\ (\bibinfo  {publisher}
  {Livraria da F\'{\i}sica},\ \bibinfo {year} {2010})\BibitemShut {NoStop}%
\bibitem [{\citenamefont {Lorenci}\ and\ \citenamefont
  {Souza}(2001)}]{DeLorenci2001417}%
  \BibitemOpen
  \bibfield  {author} {\bibinfo {author} {\bibfnamefont {V.~D.}\ \bibnamefont
  {Lorenci}}\ and\ \bibinfo {author} {\bibfnamefont {M.}~\bibnamefont
  {Souza}},\ }\href@noop {} {\bibfield  {journal} {\bibinfo  {journal} {Physics
  Letters B}\ }\textbf {\bibinfo {volume} {512}},\ \bibinfo {pages} {417 }
  (\bibinfo {year} {2001})}\BibitemShut {NoStop}%
\bibitem [{\citenamefont {Novello}\ and\ \citenamefont
  {Bittencourt}(2012)}]{Novello2012Gordonmetric}%
  \BibitemOpen
  \bibfield  {author} {\bibinfo {author} {\bibfnamefont {M.}~\bibnamefont
  {Novello}}\ and\ \bibinfo {author} {\bibfnamefont {E.}~\bibnamefont
  {Bittencourt}},\ }\href@noop {} {\bibfield  {journal} {\bibinfo  {journal}
  {Phys. Rev. D}\ }\textbf {\bibinfo {volume} {86}},\ \bibinfo {pages} {124024}
  (\bibinfo {year} {2012})}\BibitemShut {NoStop}%
\bibitem [{\citenamefont {De~Lorenci}\ \emph {et~al.}(2001)\citenamefont
  {De~Lorenci}, \citenamefont {Figueiredo}, \citenamefont {Fliche},\ and\
  \citenamefont {Novello}}]{de2001dyadosphere}%
  \BibitemOpen
  \bibfield  {author} {\bibinfo {author} {\bibfnamefont {V.~A.}\ \bibnamefont
  {De~Lorenci}}, \bibinfo {author} {\bibfnamefont {N.}~\bibnamefont
  {Figueiredo}}, \bibinfo {author} {\bibfnamefont {H.~H.}\ \bibnamefont
  {Fliche}}, \ and\ \bibinfo {author} {\bibfnamefont {M.}~\bibnamefont
  {Novello}},\ }\href@noop {} {\bibfield  {journal} {\bibinfo  {journal}
  {Astronomy and Astrophysics}\ }\textbf {\bibinfo {volume} {369}},\ \bibinfo
  {pages} {690} (\bibinfo {year} {2001})}\BibitemShut {NoStop}%
\bibitem [{\citenamefont {Bret{\'o}n}(2002)}]{breton2002geodesic}%
  \BibitemOpen
  \bibfield  {author} {\bibinfo {author} {\bibfnamefont {N.}~\bibnamefont
  {Bret{\'o}n}},\ }\href@noop {} {\bibfield  {journal} {\bibinfo  {journal}
  {Classical and Quantum Gravity}\ }\textbf {\bibinfo {volume} {19}},\ \bibinfo
  {pages} {601} (\bibinfo {year} {2002})}\BibitemShut {NoStop}%
\bibitem [{\citenamefont {Delphenich}(2006)}]{delphenich2006optanalog}%
  \BibitemOpen
  \bibfield  {author} {\bibinfo {author} {\bibfnamefont {D.~H.}\ \bibnamefont
  {Delphenich}},\ }\href@noop {} {\bibfield  {journal} {\bibinfo  {journal}
  {arXiv preprint hep-th/0610088}\ } (\bibinfo {year} {2006})}\BibitemShut
  {NoStop}%
\bibitem [{\citenamefont {Niau-Akmansoy}\ and\ \citenamefont
  {Medeiros}(2014)}]{akmansoy2013thermodynamics}%
  \BibitemOpen
  \bibfield  {author} {\bibinfo {author} {\bibfnamefont {P.}~\bibnamefont
  {Niau-Akmansoy}}\ and\ \bibinfo {author} {\bibfnamefont {L.~G.}\ \bibnamefont
  {Medeiros}},\ }\href@noop {} {\bibfield  {journal} {\bibinfo  {journal}
  {Physics Letters B}\ }\textbf {\bibinfo {volume} {738}},\ \bibinfo {pages}
  {317} (\bibinfo {year} {2014})}\BibitemShut {NoStop}%
\bibitem [{\citenamefont {De~Melo}\ \emph {et~al.}(2015)\citenamefont
  {De~Melo}, \citenamefont {Medeiros},\ and\ \citenamefont
  {Pompeia}}]{demelo2014causal}%
  \BibitemOpen
  \bibfield  {author} {\bibinfo {author} {\bibfnamefont {C.~A.~M.}\
  \bibnamefont {De~Melo}}, \bibinfo {author} {\bibfnamefont {L.~G.}\
  \bibnamefont {Medeiros}}, \ and\ \bibinfo {author} {\bibfnamefont {P.~J.}\
  \bibnamefont {Pompeia}},\ }\href@noop {} {\bibfield  {journal} {\bibinfo
  {journal} {Modern Physics Letters A}\ }\textbf {\bibinfo {volume} {30}},\
  \bibinfo {pages} {1550025} (\bibinfo {year} {2015})}\BibitemShut {NoStop}%
\bibitem [{\citenamefont {Weinberg}(1972)}]{weinberg1972gravitation}%
  \BibitemOpen
  \bibfield  {author} {\bibinfo {author} {\bibfnamefont {S.}~\bibnamefont
  {Weinberg}},\ }\href@noop {} {\emph {\bibinfo {title} {Gravitation and
  cosmology: Principle and applications of general theory of relativity}}}\
  (\bibinfo  {publisher} {John Wiley and Sons, Inc., New York},\ \bibinfo
  {year} {1972})\BibitemShut {NoStop}%
\end{thebibliography}%

\end{document}